\documentclass{article}
\usepackage[pdftex]{graphicx} % need if not making elsarticle

\usepackage{setspace}
\usepackage{amssymb,amsmath,amsthm}
\usepackage[margin=0.9in]{geometry} % reduce margins for preprint
\usepackage[]{algorithm2e}

\usepackage{verbatim} % for block comment
\usepackage{listings} % for displaying code
\usepackage{color}
\usepackage[normalem]{ulem}

\begin{document}

\title{Feature-level Malware Obfuscation in Deep Learning}

%	
%
%%% Group authors per affiliation:
%
\author{Keith Dillon}

%\cortext[mycorrespondingauthor]{Corresponding author. Tel. 1-949-478-1736}
%\ead{kdillon@newhaven.edu}
%%\author[tulaneaddress]{Yu-Ping Wang}
%
%\address[newhavenaddress]{Department of Electrical and Computer Engineering and Computer Science, \\ University of New Haven, West Haven, CT, USA}
%%\address[tulaneaddress]{Department of Biomedical Engineering, \\ Tulane University, New Orleans, LA, USA}
%
%%\address[myotheraddress]{Department of Biostatistics and Bioinformatics, Tulane University, New Orleans, LA, USA}
%	
%	

\maketitle	
		
\begin{abstract}
We consider the problem of detecting malware with deep learning models, where the malware may be combined with significant amounts of benign code. 
Examples of this include piggybacking and trojan horse attacks on a system, where malicious behavior is hidden within a useful application. 
Such added flexibility in augmenting the malware enables significantly more code obfuscation.
Hence we focus on the use of static features, particularly Intents, Permissions, and API calls, which we presume cannot be ultimately hidden from the Android system, but only augmented with yet more such features. 
We first train a deep neural network classifier  for malware classification using features of benign and malware samples. Then we demonstrate a steep increase in false negative rate (i.e., attacks succeed), simply by randomly adding features of a benign app to malware.
Finally we test the use of data augmentation to harden the classifier against such attacks. 
We find that for API calls, it is possible to reject the vast majority of attacks, where using Intents or Permissions is less successful. 
\end{abstract}

%\maketitle % make a title page that can be extracted for submission

%=====================================================================================

%
%\clearpage
%\newpage
%

\section{Introduction}

A number of technologies are available to attackers which can make the automated detection of malware difficult. 
Code obfuscation techniques can be used to circumvent conventional signature-based detection approaches. 
This includes simple techniques such as encryption methods for protecting intellectual property \cite{wermke_large_2018},  as well as increasingly-sophisticated ways to change code without altering its function \cite{you_malware_2010}. 
The rapid appearance and sheer volume of new malware variants has prompted the use of machine learning approaches to keep up \cite{rieck_automatic_2011}.
Initial machine learning approaches utilized expert feature engineering, by first using tools to extract important features.
Then shallow classification methods such as support vector machines were applied to distinguish malware from benign apps, and various types of static or dynamic features were compared \cite{wang_constructing_2019}.

More recent research has utilized deep learning, which is able to learn important features from large and unstructured datasets \cite{mclaughlin_deep_2017,cakir_malware_2018,pascanu_malware_2015,le_deep_2018,xu_deeprefiner_2018}. 
Deep Learning models have been applied to traditional static and/or dynamic features \cite{grosse_adversarial_2016,grosse_adversarial_2017}.
Deep networks are also applied directly on the bytecode or decompiled code \cite{xu_deeprefiner_2018,raff_malware_2017}, utilizing techniques for processing sequences such as from natural language processing  \cite{cakir_malware_2018,pascanu_malware_2015}, or convolutional neural nets \cite{mclaughlin_deep_2017,hasegawa_one-dimensional_2018,cui_detection_2018}, which has also been prominently applied to sequences and text.
Typically, these architectures exact features based on short sequences of instructions, making them again vulnerable to sufficiently-sophisticated obfuscation techniques which reorder code into new sequences.

A related area of deep learning research involves adversarial examples \cite{szegedy_intriguing_2014}.
Most famously, this entails a imperceptibly-small perturbation of an image which causes an otherwise accurate deep neural network to classify it incorrectly.
Adversarial example research has been extended to other data types \cite{xu_adversarial_2019}, including malware \cite{suciu_exploring_2019}.
For malware, the problem is essentially the same as code obfuscation under the constraint that the adjustment to the malware file must be kept small and not affect functionality \cite{kolosnjaji_adversarial_2018}. 
This has been demonstrated with classification based on static or dynamic features \cite{grosse_adversarial_2016,grosse_adversarial_2017}, as well as sequence data \cite{suciu_exploring_2019}.
There has also been significant research on hardening deep learning against adversarial attacks \cite{podschwadt_effectiveness_2019,grosse_adversarial_2016,grosse_adversarial_2017,xu_feature_2018,fleshman_non-negative_2019}.
One of the most prominent approaches is a form of data augmentation called ``adversarial training", where the network is simply trained using correctly-labeled adversarial malware samples. 
Nonetheless, the attacks used in adversarial example research are not necessarily realistic \cite{gilmer_motivating_2018}, due to the artificial constraints on perturbation-size. 

In this paper we will address malware obfuscation involving large alternations to both size and function, by allowing the addition of benign features to the malware.  
First we train a dense network to classify malware.
Then we demonstrate the effect of obfuscation by randomly adding features from benign samples, which causes a large increase in false negative rate (i.e. the obfuscation succeeds).
Lastly we demonstrate the improvement gained by training with obfuscated samples, where the original performance can be nearly achieved with the proper choice of feature types.

%
%
%
%CNN malware AAs \cite{suciu exploring}
%
%detecting AAs in malware \cite{Xu feature}
%
%weight-constraints for hardening \cite{fleshman non-neg}
%
%randomly nullifying features for hardening  \cite{wang adverrsary}
%
%criticism of AA research as unrealistic attacks \cite{gilmer motivating}
%
%general AA attacks and defense review for malware \cite{powdschwadt on-effectiveness}
%
%

\section{Results}

\subsection{Data Collection and Preprocessing}

We used the AndroParse dataset \cite{schmicker_androparse_2019} consisting of approximately 90k Apps, 60k benign and 30k malicious. 
AndroParse extracts a variety of features including API's, Intents, Permissions, which we used in this research.
%
%Details about the data can be found...
%
%The features were collected using...
%
The dataset contained 3438 unique intents and 19,827 unique permissions which were used as features.
A set of API calls was formed by extracting all function calls, then removing the function name and recursively truncating the class hierarchy (e.g., \verb|com.google.android.gms.internal.zztn.zzbS| was broken into \verb|com.google.android.gms.internal.zztn|, \verb|com.google.android.gms.internal|, \verb|com.google.android.gms|, and \verb|com.google.android|). 
Finally the most common 20k elements were chosen from this set to use as features.
After discarding a small number of samples for which there were zero features, we ultimately had 89,678 samples with 61,180 benign and 28,498 malicious.

The samples were randomly shuffled, then 30 percent were selected for the test set and 70 percent for the training set.
As we did not perform early-stopping or extensive parameter tuning (see next section), the test set was also used as the validation set during training to demonstrate convergence.

\subsection{Network Architecture}

We ultimately used a network consisting of 20 dense layers with 1024 nodes each, plus a single-node output layer to form the binary classifier. 
We did not experience any noticeable overfitting or a need for careful hyperparameter selection. So we simply chose a dense architecture with standard optimization settings, finding essentially the same results as for variations on the architecture.
The network was implemented in Keras \cite{chollet_deep_2017} (version 2.2.4-tf) with TensorFlow backend (verson 2.0.0) on a NVidia 2080ti GPU. 
The model summary generated by Keras is given in the appendix for the case when all feature types were used.
As noted above, we considered various architectures formed via a stack of deep layers. 
While it was evident that a deep network performed better than a shallow network, doubling or halving the depth of the deep network did not affect results noticeably.
This was despite the fact that there were far more trainable parameters than training samples for the baseline test, suggesting that most of the samples were easy to discriminate. 
A relatively complex network was chosen to allow for additional adaptability in the subsequent training with obfuscated data, where there would effectively be unlimited training samples due to data augmentation.

Typical network settings were used: binary cross-entropy loss, rectified linear unit (ReLU) activation on hidden layers, sigmoid activation on final layer, stochastic gradient descent (learning rate 0.1) with Nesterov momentum (rate = 0.9) and decay of $10^{-6}$.
A batch size of 2048 was used for speed as the data transfer to the GPU was the bottleneck with this network. Different batch sizes had no noticeable effect classifier performance. 

\subsection{Baseline Testing}

\begin{figure}[h!] \centering % % 0.25 spacing
	\scalebox{0.65}{\includegraphics[trim=0.0in -.1in 0.0in 0.0in]{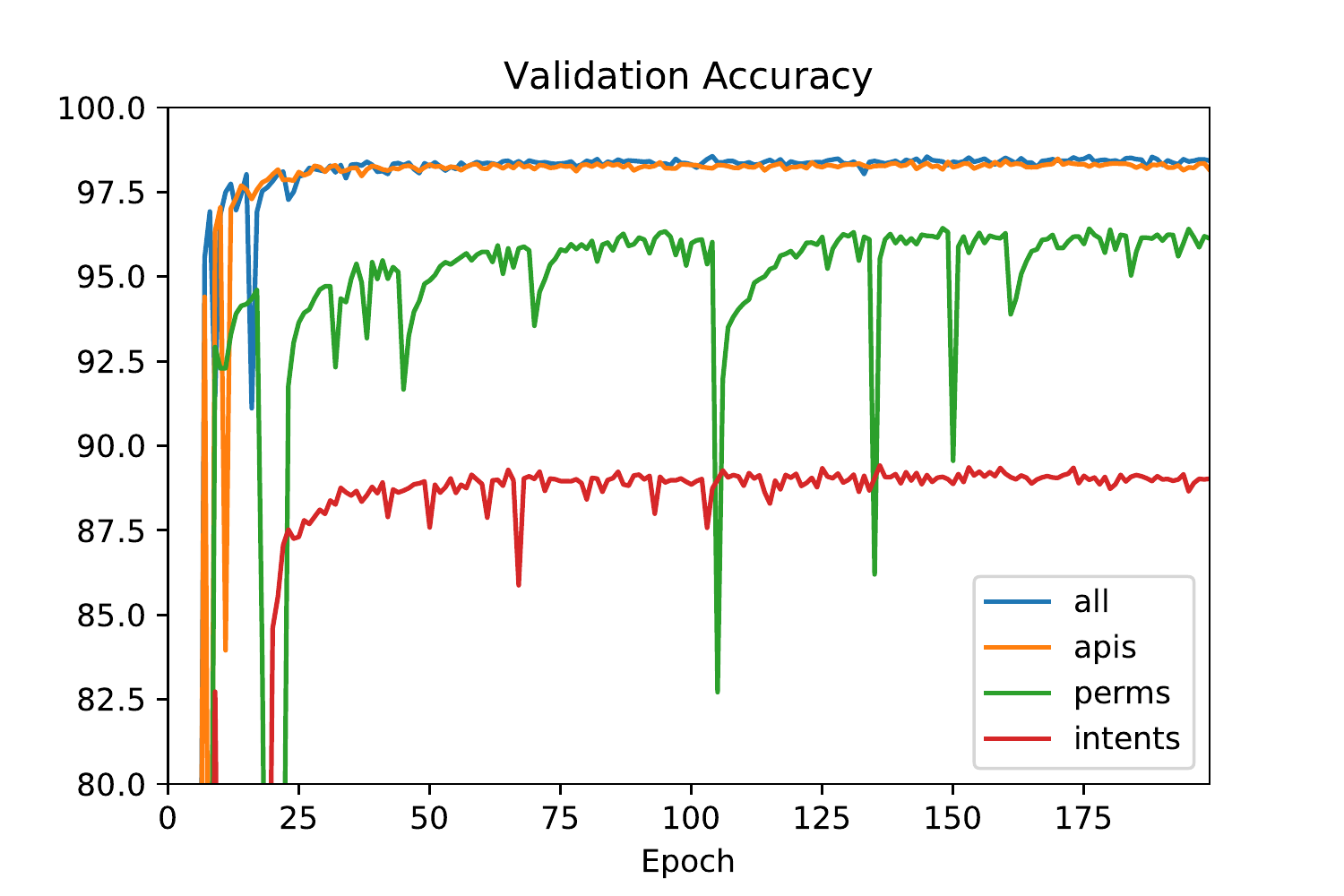}} % lbrt 
	\caption{Validation accuracy (percentage) during training for baseline model using different features.} 
	\label{fig_histb}
\end{figure}

\begin{figure}[h!] \centering % % 0.25 spacing
	\scalebox{0.7}{\includegraphics[trim=0.0in -.1in 0.0in 0.0in]{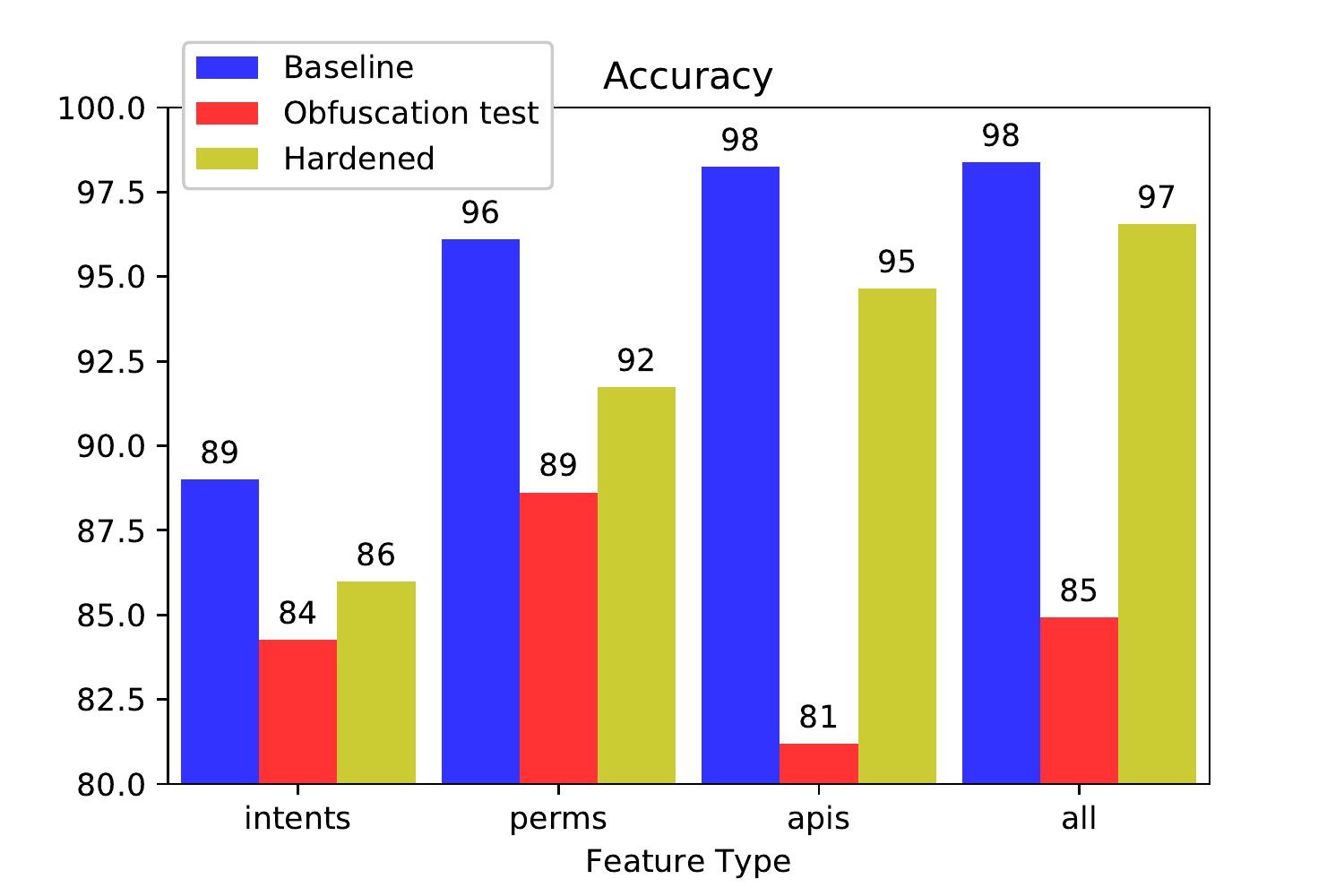}} % lbrt 
	\caption{Accuracy percentage comparing baseline mode trained and tested on baseline datasets (blue), baseline model trained on baseline data and tested on  obfuscated data (green), model trained and tested on obfuscated data (red).} 
	\label{fig_dnnacc}
\end{figure}

\begin{figure}[h!] \centering % % 0.25 spacing
	\scalebox{0.7}{\includegraphics[trim=0.0in -.1in 0.0in 0.0in]{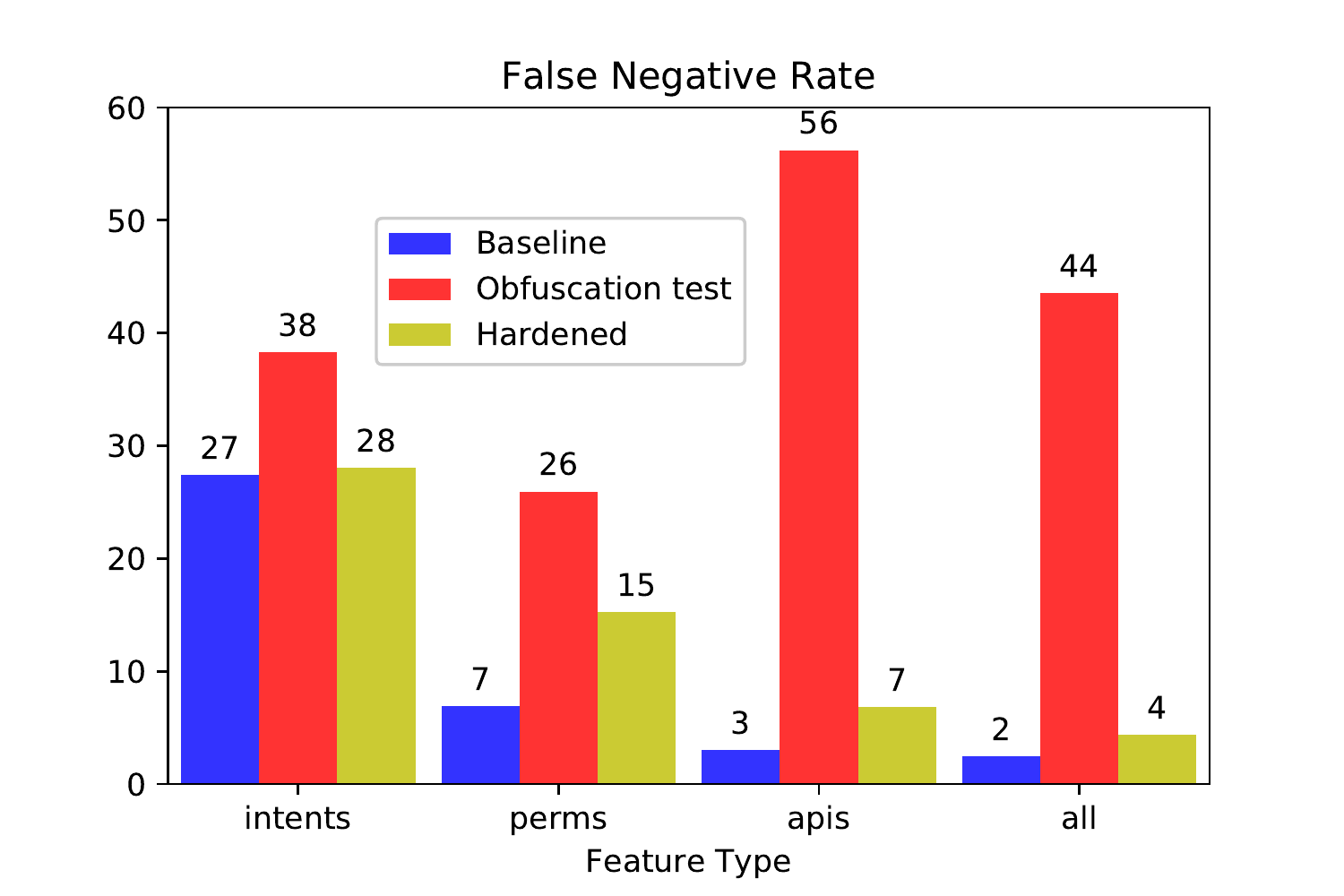}} % lbrt 
	\caption{False negative rate percentage (malware incorrectly classified as benign) comparing baseline mode trained and tested on baseline datasets (blue), baseline model trained on baseline data and tested on  obfuscated data (green), model trained and tested on obfuscated data (red).} 
	\label{fig_dnnfnr}
\end{figure}

\begin{figure}[h!] \centering % % 0.25 spacing
	\scalebox{0.7}{\includegraphics[trim=0.0in -.1in 0.0in 0.0in]{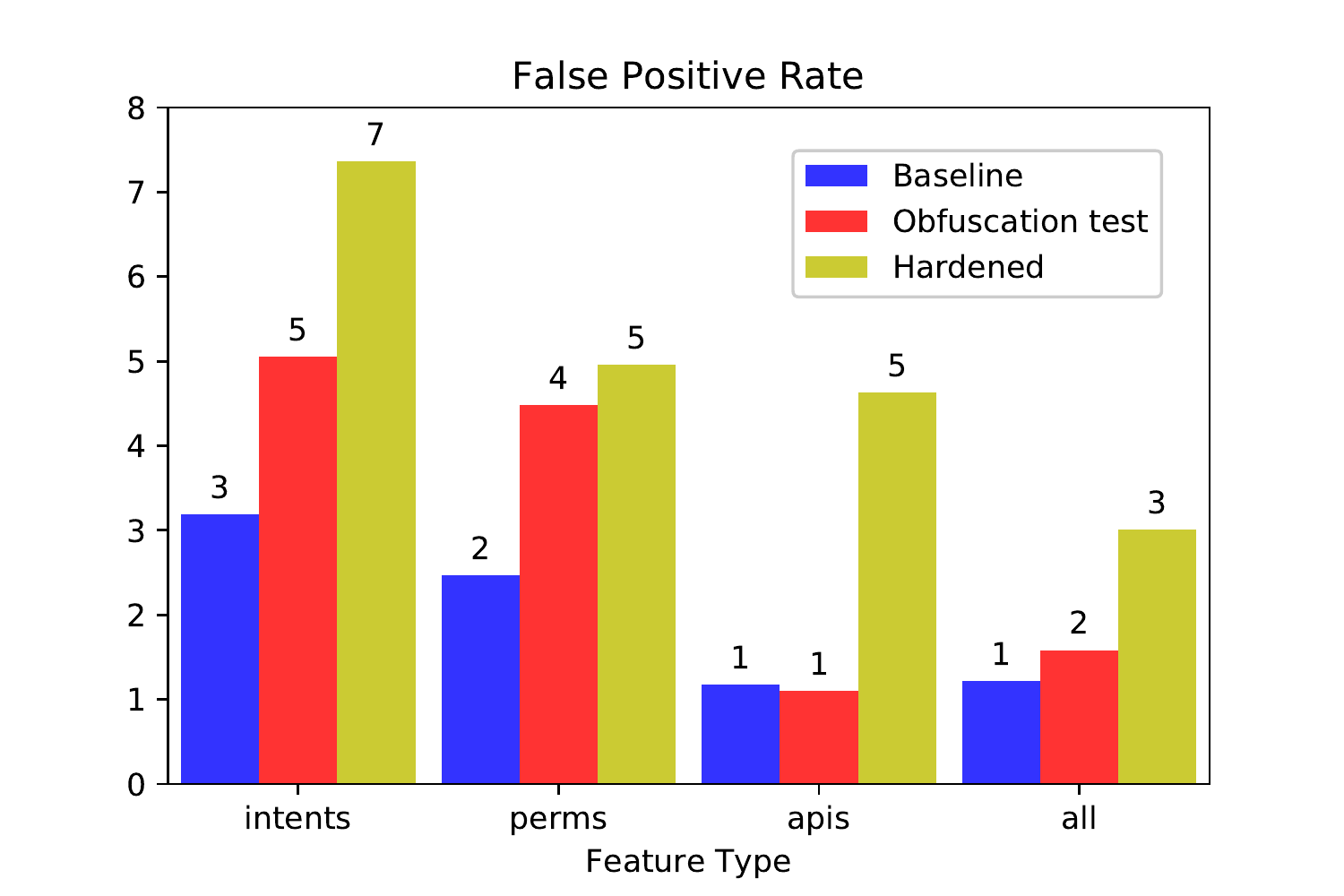}} % lbrt 
	\caption{False positive rate percentage (benign incorrectly classified as malware) comparing baseline mode trained and tested on baseline datasets (blue), baseline model trained on baseline data and tested on obfuscated data (green), model trained and tested on obfuscated data (red).} 
	\label{fig_dnnfpr}
\end{figure}

The baseline performance in this case refers to the network trained and tested without any obfuscation, i.e., a conventional application of machine learning to the dataset described previously. 
This provides an upper limit on performance we would expect compared to cases where the malicious samples are subsequently obfuscated to hide their true class from the classifier. 
The accuracy of this network during training in plotted in Fig. \ref{fig_histb} using different feature-types  independently, and compared to using all three feature-types.
The training set was randomly-shuffled after each epoch, which may be responsible for the noisy convergence initially. 
The final test accuracy, false-negative rate (FNR) and false-positive rate (FPR) are given in Figs. \ref{fig_dnnacc}, \ref{fig_dnnfnr}, and \ref{fig_dnnfpr}, respectively. 
We find that using API's resulted in superior performance compared to intents and permissions, with a very small improvement achieved by combining all three feature types.

\subsection{Obfuscation Testing}
To test the baseline network (as well as hardened networks we will describe in the next section) with obfuscated samples, we formed new test set by randomly adding features from benign samples in the test set to the malicious samples in the test set.
We compared using benign features from the training set instead of the test set, and found those from the test set to result in marginally worse performance.  
The final obfuscated test set was the same size as the original test set.

The obfuscated test accuracy, FNR, and FPR are also given in Figs. \ref{fig_dnnacc}, \ref{fig_dnnfnr}, and \ref{fig_dnnfpr}, respectively. 
We found a moderate reduction in accuracy (around 15 percent) and slightly-worse FPR (one or two percent), but a very large increase in FNR (i.e., malware which has been successfully obfuscated).
This is especially bad for the previously-best-performing features, API's, where the FNR increased from three percent to 55 percent.

\subsection{Network Hardening}
\begin{figure}[h!] \centering % % 0.25 spacing
	\scalebox{0.65}{\includegraphics[trim=0.0in -.1in 0.0in 0.0in]{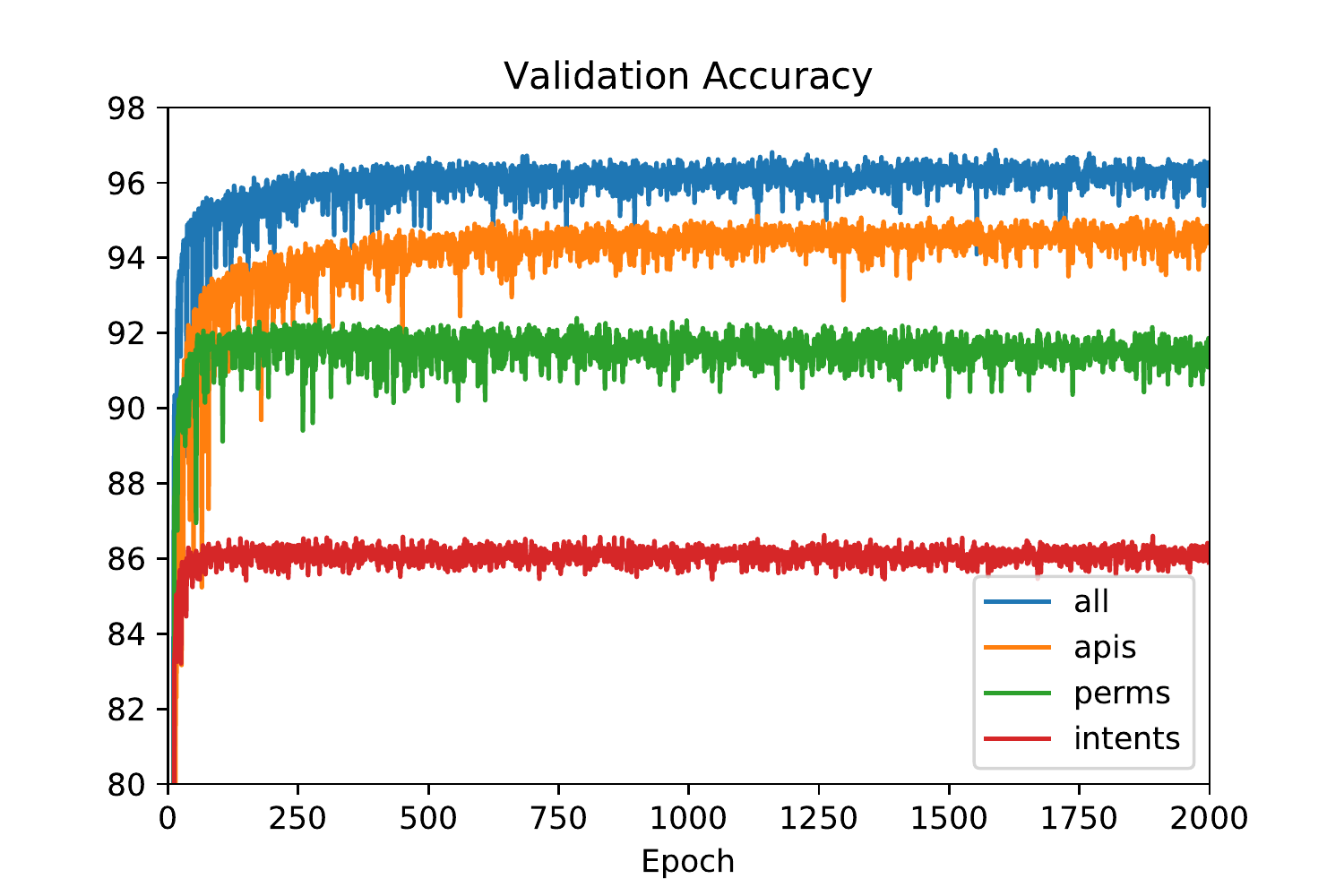}} % lbrt 
	\caption{Validation accuracy during training with hardening for different datasets. Note that the test set is not obfuscated here.} 
	\label{fig_histo}
\end{figure}

In order to harden the network against obfuscated features, we used data augmentation \cite{chollet_deep_2017}.
We created a data generator which randomly produces new batches of data, by first selecting a batch of training data, then adding features from randomly-selected benign samples to the malicious samples.   
Each new epoch used new random obfuscating benign features as it cycled through batches of the training set.
The training set was also randomly shuffled after each epoch.
The validation accuracy  during training (note the validation data was not obfuscated for this calculation) is plotted in Fig. \ref{fig_histo} using different feature-types  independently, and compared to using all three feature-types.
As the training data was different each epoch, we tested the ability to program a degree invariance to benign features by running for many more epochs.
However after the initial increase inaccuracy, the subsequent improvement for many more epochs was very small.
Further, this time we noted a larger improvement in combining all feature types, versus simply using API's.

The hardened test accuracy, FNR, and FPR using an obfuscated test set are given in Figs. \ref{fig_dnnacc}, \ref{fig_dnnfnr}, and \ref{fig_dnnfpr}, respectively. 
We found that when using all feature types we are almost able to achieve comparable performance to the baseline network on non-obfuscated data. 
We again found that intents and permissions performed significantly worse when used alone as features.

\section{Discussion}

We considered the open and daunting problem of detecting malware subject to large (obfuscating) changes, i.e., attacks based on adding new functionality and additional code to malware, or, in the extreme, adding malware to benign applications. 
We presumed that we would nonetheless be able to extract the features of such obfuscated malware, and so tested the ability to classify and train using augmented sets of features. 
We found both that it is very easy to hide malware from a conventional deep network using this obfuscation, but also that hardening the network against such an attack should work well.
The choice of features was a very important factor, with API's (or API's plus additional feature types) outperforming intents or permissions. 

The data augmentation approach leads to an ``invariance programming'' perspective, where the goal is to generate samples which program the network to ignore certain latent variables.
In this case the latent variable is the amount of benign features, so our network performs as an asymmetrical detector, only indicating presence or absence of malicious features. 
Of course there will remain some overlap between benign and malicious functionality, which may inform decisions on how to better secure systems by restricting functionality.

A number of new directions remain for further research.
It would be useful explore the different performance for different malware families. 
It would also be very valuable to be able to skip the feature extraction stage entirely and operate directly on raw bytecode or executables. 
This would produce a significant increase in dimensionality of the problem, requiring a great deal more data (including a much more sophisticated data generator).

\section*{Appendix: Dense model summary}

\begin{lstlisting}
Model: "sequential"
_________________________________________________________________
Layer (type)                 Output Shape              Param #   
=================================================================
dense (Dense)                (None, 1024)              44288000  
_________________________________________________________________
dense_1 (Dense)              (None, 1024)              1049600   
_________________________________________________________________
dense_2 (Dense)              (None, 1024)              1049600   
_________________________________________________________________
dense_3 (Dense)              (None, 1024)              1049600   
_________________________________________________________________
dense_4 (Dense)              (None, 1024)              1049600   
_________________________________________________________________
dense_5 (Dense)              (None, 1024)              1049600   
_________________________________________________________________
dense_6 (Dense)              (None, 1024)              1049600   
_________________________________________________________________
dense_7 (Dense)              (None, 1024)              1049600   
_________________________________________________________________
dense_8 (Dense)              (None, 1024)              1049600   
_________________________________________________________________
dense_9 (Dense)              (None, 1024)              1049600   
_________________________________________________________________
dense_10 (Dense)             (None, 1024)              1049600   
_________________________________________________________________
dense_11 (Dense)             (None, 1024)              1049600   
_________________________________________________________________
dense_12 (Dense)             (None, 1024)              1049600   
_________________________________________________________________
dense_13 (Dense)             (None, 1024)              1049600   
_________________________________________________________________
dense_14 (Dense)             (None, 1024)              1049600   
_________________________________________________________________
dense_15 (Dense)             (None, 1024)              1049600   
_________________________________________________________________
dense_16 (Dense)             (None, 1024)              1049600   
_________________________________________________________________
dense_17 (Dense)             (None, 1024)              1049600   
_________________________________________________________________
dense_18 (Dense)             (None, 1024)              1049600   
_________________________________________________________________
dense_19 (Dense)             (None, 1024)              1049600   
_________________________________________________________________
dense_20 (Dense)             (None, 1)                 1025      
=================================================================
Total params: 64,231,425
Trainable params: 64,231,425
Non-trainable params: 0
_________________________________________________________________
\end{lstlisting}

%
%features hidden from static analysis - reflection
%
%

%\section*{References} % need if using elsarticle
\bibliography{../../zoterorefs}

\begin{thebibliography}{10}

\bibitem{cakir_malware_2018}
Bugra Cakir and Erdogan Dogdu.
\newblock Malware classification using deep learning methods.
\newblock In {\em Proceedings of the {ACMSE} 2018 {Conference}}, {ACMSE} '18,
  pages 1--5, Richmond, Kentucky, March 2018. Association for Computing
  Machinery.

\bibitem{chollet_deep_2017}
François Chollet.
\newblock {\em Deep {Learning} with {Python}}.
\newblock Manning Publications Company, October 2017.
\newblock Google-Books-ID: Yo3CAQAACAAJ.

\bibitem{cui_detection_2018}
Zhihua Cui, Fei Xue, Xingjuan Cai, Yang Cao, Gai-ge Wang, and Jinjun Chen.
\newblock Detection of {Malicious} {Code} {Variants} {Based} on {Deep}
  {Learning}.
\newblock {\em IEEE Transactions on Industrial Informatics}, 14(7):3187--3196,
  July 2018.

\bibitem{fleshman_non-negative_2019}
William Fleshman, Edward Raff, Jared Sylvester, Steven Forsyth, and Mark
  McLean.
\newblock Non-{Negative} {Networks} {Against} {Adversarial} {Attacks}.
\newblock {\em arXiv:1806.06108 [cs, stat]}, January 2019.
\newblock arXiv: 1806.06108.

\bibitem{gilmer_motivating_2018}
Justin Gilmer, Ryan~P. Adams, Ian Goodfellow, David Andersen, and George~E.
  Dahl.
\newblock Motivating the {Rules} of the {Game} for {Adversarial} {Example}
  {Research}.
\newblock {\em arXiv:1807.06732 [cs, stat]}, July 2018.
\newblock arXiv: 1807.06732.

\bibitem{grosse_adversarial_2016}
Kathrin Grosse, Nicolas Papernot, Praveen Manoharan, Michael Backes, and
  Patrick McDaniel.
\newblock Adversarial {Perturbations} {Against} {Deep} {Neural} {Networks} for
  {Malware} {Classification}.
\newblock {\em arXiv:1606.04435 [cs]}, June 2016.
\newblock arXiv: 1606.04435.

\bibitem{grosse_adversarial_2017}
Kathrin Grosse, Nicolas Papernot, Praveen Manoharan, Michael Backes, and
  Patrick McDaniel.
\newblock Adversarial {Examples} for {Malware} {Detection}.
\newblock In Simon~N. Foley, Dieter Gollmann, and Einar Snekkenes, editors,
  {\em Computer {Security} – {ESORICS} 2017}, Lecture {Notes} in {Computer}
  {Science}, pages 62--79, Cham, 2017. Springer International Publishing.

\bibitem{hasegawa_one-dimensional_2018}
Chihiro Hasegawa and Hitoshi Iyatomi.
\newblock One-dimensional convolutional neural networks for {Android} malware
  detection.
\newblock In {\em 2018 {IEEE} 14th {International} {Colloquium} on {Signal}
  {Processing} {Its} {Applications} ({CSPA})}, pages 99--102, March 2018.
\newblock ISSN: null.

\bibitem{kolosnjaji_adversarial_2018}
Bojan Kolosnjaji, Ambra Demontis, Battista Biggio, Davide Maiorca, Giorgio
  Giacinto, Claudia Eckert, and Fabio Roli.
\newblock Adversarial {Malware} {Binaries}: {Evading} {Deep} {Learning} for
  {Malware} {Detection} in {Executables}.
\newblock In {\em 2018 26th {European} {Signal} {Processing} {Conference}
  ({EUSIPCO})}, pages 533--537, September 2018.
\newblock ISSN: 2219-5491.

\bibitem{le_deep_2018}
Quan Le, Oisín Boydell, Brian Mac~Namee, and Mark Scanlon.
\newblock Deep learning at the shallow end: {Malware} classification for
  non-domain experts.
\newblock {\em Digital Investigation}, 26:S118--S126, July 2018.

\bibitem{mclaughlin_deep_2017}
Niall McLaughlin, Jesus Martinez~del Rincon, BooJoong Kang, Suleiman Yerima,
  Paul Miller, Sakir Sezer, Yeganeh Safaei, Erik Trickel, Ziming Zhao, Adam
  Doupé, and Gail Joon~Ahn.
\newblock Deep {Android} {Malware} {Detection}.
\newblock In {\em Proceedings of the {Seventh} {ACM} on {Conference} on {Data}
  and {Application} {Security} and {Privacy}}, {CODASPY} '17, pages 301--308,
  Scottsdale, Arizona, USA, March 2017. Association for Computing Machinery.

\bibitem{pascanu_malware_2015}
R.~Pascanu, J.~W. Stokes, H.~Sanossian, M.~Marinescu, and A.~Thomas.
\newblock Malware classification with recurrent networks.
\newblock In {\em 2015 {IEEE} {International} {Conference} on {Acoustics},
  {Speech} and {Signal} {Processing} ({ICASSP})}, pages 1916--1920, April 2015.

\bibitem{podschwadt_effectiveness_2019}
Robert Podschwadt and Hassan Takabi.
\newblock On {Effectiveness} of {Adversarial} {Examples} and {Defenses} for
  {Malware} {Classification}.
\newblock In Songqing Chen, Kim-Kwang~Raymond Choo, Xinwen Fu, Wenjing Lou, and
  Aziz Mohaisen, editors, {\em Security and {Privacy} in {Communication}
  {Networks}}, Lecture {Notes} of the {Institute} for {Computer} {Sciences},
  {Social} {Informatics} and {Telecommunications} {Engineering}, pages
  380--393, Cham, 2019. Springer International Publishing.

\bibitem{raff_malware_2017}
Edward Raff, Jon Barker, Jared Sylvester, Robert Brandon, Bryan Catanzaro, and
  Charles Nicholas.
\newblock Malware {Detection} by {Eating} a {Whole} {EXE}.
\newblock {\em arXiv:1710.09435 [cs, stat]}, October 2017.
\newblock arXiv: 1710.09435.

\bibitem{rieck_automatic_2011}
Konrad Rieck, Philipp Trinius, Carsten Willems, and Thorsten Holz.
\newblock Automatic analysis of malware behavior using machine learning.
\newblock {\em Journal of Computer Security}, 19(4):639--668, January 2011.

\bibitem{schmicker_androparse_2019}
Robert Schmicker, Frank Breitinger, and Ibrahim Baggili.
\newblock {AndroParse} - {An} {Android} {Feature} {Extraction} {Framework} and
  {Dataset}.
\newblock In Frank Breitinger and Ibrahim Baggili, editors, {\em Digital
  {Forensics} and {Cyber} {Crime}}, Lecture {Notes} of the {Institute} for
  {Computer} {Sciences}, {Social} {Informatics} and {Telecommunications}
  {Engineering}, pages 66--88, Cham, 2019. Springer International Publishing.

\bibitem{suciu_exploring_2019}
Octavian Suciu, Scott~E. Coull, and Jeffrey Johns.
\newblock Exploring {Adversarial} {Examples} in {Malware} {Detection}.
\newblock {\em arXiv:1810.08280 [cs, stat]}, April 2019.
\newblock arXiv: 1810.08280.

\bibitem{szegedy_intriguing_2014}
Christian Szegedy, Wojciech Zaremba, Ilya Sutskever, Joan Bruna, Dumitru Erhan,
  Ian Goodfellow, and Rob Fergus.
\newblock Intriguing properties of neural networks.
\newblock {\em arXiv:1312.6199 [cs]}, February 2014.
\newblock arXiv: 1312.6199.

\bibitem{wang_constructing_2019}
Wei Wang, Meichen Zhao, Zhenzhen Gao, Guangquan Xu, Hequn Xian, Yuanyuan Li,
  and Xiangliang Zhang.
\newblock Constructing {Features} for {Detecting} {Android} {Malicious}
  {Applications}: {Issues}, {Taxonomy} and {Directions}.
\newblock {\em IEEE Access}, 7:67602--67631, 2019.

\bibitem{wermke_large_2018}
Dominik Wermke, Nicolas Huaman, Yasemin Acar, Bradley Reaves, Patrick Traynor,
  and Sascha Fahl.
\newblock A {Large} {Scale} {Investigation} of {Obfuscation} {Use} in {Google}
  {Play}.
\newblock In {\em Proceedings of the 34th {Annual} {Computer} {Security}
  {Applications} {Conference}}, {ACSAC} '18, pages 222--235, San Juan, PR, USA,
  December 2018. Association for Computing Machinery.

\bibitem{xu_adversarial_2019}
Han Xu, Yao Ma, Haochen Liu, Debayan Deb, Hui Liu, Jiliang Tang, and Anil~K.
  Jain.
\newblock Adversarial {Attacks} and {Defenses} in {Images}, {Graphs} and
  {Text}: {A} {Review}.
\newblock {\em arXiv:1909.08072 [cs, stat]}, October 2019.
\newblock arXiv: 1909.08072.

\bibitem{xu_deeprefiner_2018}
Ke~Xu, Yingjiu Li, Robert~H. Deng, and Kai Chen.
\newblock {DeepRefiner}: {Multi}-layer {Android} {Malware} {Detection} {System}
  {Applying} {Deep} {Neural} {Networks}.
\newblock In {\em 2018 {IEEE} {European} {Symposium} on {Security} and
  {Privacy} ({EuroS} {P})}, pages 473--487, April 2018.
\newblock ISSN: null.

\bibitem{xu_feature_2018}
Weilin Xu, David Evans, and Yanjun Qi.
\newblock Feature {Squeezing}: {Detecting} {Adversarial} {Examples} in {Deep}
  {Neural} {Networks}.
\newblock {\em Proceedings 2018 Network and Distributed System Security
  Symposium}, 2018.
\newblock arXiv: 1704.01155.

\bibitem{you_malware_2010}
Ilsun You and Kangbin Yim.
\newblock Malware {Obfuscation} {Techniques}: {A} {Brief} {Survey}.
\newblock In {\em 2010 {International} {Conference} on {Broadband}, {Wireless}
  {Computing}, {Communication} and {Applications}}, pages 297--300, November
  2010.
\newblock ISSN: null.

\end{thebibliography}
\bibliographystyle{plain}

\end{document}